\documentclass[review]{elsarticle}

\usepackage{lineno,hyperref}

\journal{Journal of \LaTeX\ Templates}

\usepackage{doi}
\usepackage{hyperref}
\hypersetup{
	colorlinks=true,        
	linkcolor=blue,         
	citecolor=cyan,         
}

\usepackage{graphicx}
\usepackage{dcolumn}
\usepackage{bm}
\usepackage{color}

\usepackage{amsmath}
\usepackage{amssymb}

\begin{document}

\begin{frontmatter}

\title{Charged particle and epicyclic motions around $4D$ Einstein-Gauss-Bonnet \\black hole immersed in an external magnetic field 
}

\author[af1,af2,af3,af4]{Sanjar Shaymatov}
\ead{sanjar@astrin.uz}

\author[af5]{Jaroslav Vrba}
\ead{jaroslav.vrba@physics.slu.cz}

\author[af6]{Daniele Malafarina}
\ead{daniele.malafarina@nu.edu.kz}

\author[af2,af3,af4]{Bobomurat Ahmedov}
\ead{ahmedov@astrin.uz}

\author[af5]{Zden\v{e}k Stuchl\'{i}k}
\ead{zdenek.stuchlik@physics.slu.cz}

\address[af1]{Institute for Theoretical Physics and Cosmology, Zheijiang University of Technology, Hangzhou 310023, China}
\address[af2]{Ulugh Beg Astronomical Institute, Astronomy St. 33, Tashkent 100052, Uzbekistan}
\address[af3]{National University of Uzbekistan, Tashkent 100174, Uzbekistan}
\address[af4]{Tashkent Institute of Irrigation and Agricultural Mechanization Engineers,\\ Kori Niyoziy 39, Tashkent 100000, Uzbekistan}
\address[af5]{Research Centre for Theoretical Physics and Astrophysics, Institute of Physics in Opava, Silesian University in Opava,
	Bezru\v{c}ovo n\'{a}m\v{e}st\'{i} 13, CZ-74601 Opava, Czech Republic}
\address[af6]{Department of Physics, Nazarbayev University, Kabanbay Batyr 53, 010000 Nur-Sultan, Kazakhstan}

\begin{abstract}
We investigate particle motion in the vicinity of a $4D$ Einstein-Gauss-Bonnet (EGB) black hole immersed in external asymptotically uniform magnetic field. It is well known that magnetic fields can strongly affect charged particle motion in the black hole vicinity due to the Lorenz force. We find that the presence of the Gauss-Bonnet (GB) coupling gives rise to a similar effect, reducing the radius of the innermost stable circular orbit (ISCO) with respect to the purely relativistic Schwarzschild black hole. 
Further, we consider particle collisions in the black hole vicinity to determine the center of mass energy and show that this energy increases with respect to the Schwarzschild case due to the effect of the GB term. 
Finally, we consider epicyclic motion and its frequencies and resonance as a mean to test the predictions of the model against astrophysical observations. In particular we test which values of the parameters of the theory best fit the 3:2 resonance of high-frequency quasi-periodic oscillations in three low-mass X-ray binaries.

\end{abstract}

\end{frontmatter}


\section{Introduction}
\label{introduction}

Among other reasons, the fascination for black holes comes from the fact that they are described by very simple mathematical equations that are used to model existing astrophysical phenomena which allow us to explore the limits of the relativistic theory.
Until now the properties of gravitational waves detected by the LIGO and Virgo scientific collaborations \cite{Abbott16a,Abbott16b} have not shown departures from the expected behaviour of general relativistic black holes. However, we know that General Relativity (GR) is an incomplete theory and it is possible that gravity in the strong field is better described by some other theory of which GR is the low energy limit. In this context higher order theories have been considered as possible extensions of GR~\cite{Dadhich12c}.
While Einstein gravity is constructed from linear order in Riemann curvature,  Gauss-Bonnet (GB) gravity is a quadratic order theory which contains higher order invariants, thus belonging to the class of Lovelock theories~\cite{Lovelock1971}. 
While Lovelock theories are a generalization of Einstein's theory that is valid in arbitrary $D$ dimensions it was generally believed that GB gravity would give non vanishing contribution only in $D>4$, a result known as Lovelock theorem.
However, it was recently suggested that a 4-dimensional Einstein-Gauss-Bonnet ($4D$ EGB) theory could exist and that Lovelock's theorem could be bypassed by a suitable redefinition of the GB coupling constant~\cite{Glavan20prl}.
The new $4D$ EGB is presently under scrutiny on the basis of two main arguments. One involving the ill-posedness of the action for the theory~\cite{Gurses20egb,Mahapatra20egb} and the other regarding the validity of the rescaling of the GB constant, that may be possible only for systems with certain symmetries~\cite{Hennigar20egb,Arrechea20egb}. 
Both objections, if valid, may invalidate the $4D$ EGB theory as an alternative to Einstein's theory. However, the fact remains that solutions with high symmetries exist, they have clear physical interpretation that mirrors the corresponding solutions in GR, and may be regarded as coming from an effective prescription for the lower dimensional limit of GB gravity.
{ In particular, the theory admits black hole solutions that, while mimicking the GR behaviour at large distances exhibit striking differences from their Einstein counterparts in the strong field and near horizon regions. For example, in \cite{Dadhich20egb} it was shown that the causal structure of the 4D EGB black hole deviates from the general relativistic case in the vicinity of the singularity which becomes time-like for the 4D EGB black hole, while it is space-like in GR.}
For these reasons, the investigation of the properties of such solutions has value regardless of the validity of the theory and has attracted great interest~\cite[see, e.g.][]{Liu20egb,Guo20egb,Wei20egb,Kumar20egb,Konoplya20egb,
Churilova20egb,Malafarina20egb,Aragon20egb,Mansoori20egb,Ge20egb,Rayimbaev20egb,Chakraborty20egb}. At the same time, validity of the 4D EGB theory possibly with some modifications has been advocated in new theoretical frameworks that may help to overcome the above mentioned shortcomings (see for example~\cite{Odintsov20egb,Odintsov20plb,Lin20egb,Aoki20egb}).  

In the present paper, we consider high energy phenomena such as collision processes in the background geometry of a $4D$ EGB black hole immersed in an external asymptotically uniform magnetic field and the epicyclic motion around stable circular orbits and its applications to the quasi-periodic oscillations (QPOs) observed in microquasars.

In recent years much attention has been devoted to the study of charged particle motion in the background of black holes immersed in an external magnetic field in GR and alternative theories of gravity~\cite{Prasanna80,Kovar08,Kovar10,
	Frolov10,Aliev02,Abdujabbarov10,Shaymatov14,
	Kolos15,Stuchlik16,Tursunov16,Shaymatov15,shaymatov19b,Narzilloev19,Pavlovic19,Shaymatov20b,Haydarov20,Haydarov20b,Narzilloev20a}. 
Much of the analysis in the framework of particle motion has been motivated by probing the nature of winds and jets coming from active galactic nuclei (AGN)~\cite{Fender04mnrs,Auchettl17ApJ,IceCube17b}. 
To explain such phenomena a number of energetic mechanisms have been proposed, and one of the most significant is the process of high energy particle collisions occurring in the near horizon region. The original model was  proposed by Banados, Silk and West (BSW) \cite{banados09} and a large amount of work has been done since then to test the BSW effect in various frameworks, ~\cite[see,
e.g.][]{Grib11,Jacobson10,Harada11b,Wei10,Zaslavskii10,
	Zaslavskii11b,Zaslavskii11c,Kimura11,Banados11,Abdujabbarov15a,Frolov12,Liu11,Atamurotov13a,Stuchlik11a,
	Stuchlik12a,Igata12,Stuchlik13,Shaymatov13,Tursunov13,Shaymatov18a,Stuchlik20}.
Furthermore, the BSW thought experiment has been extended to the context of alternative theories of gravity~\cite{Stuchlik14a} as well as to solutions with naked singularities~\cite{Patil10,Patil11,Patil11b}. 
Particle collisions are a useful tool to probe the black hole geometry and may describe processes occurring in the environment of astrophysical black holes 
~\cite{Blandford1977,Wagh89,Morozova14,Alic12ApJ,Moesta12ApJ} such as jets from active galactic nuclei~\cite{McKinney07}. 
Arguably, the BSW model and the Penrose process~\cite{Penrose69} 
are the most studied mechanisms for the production of high energy particles in the vicinity of black holes~Refs.~\cite{Dadhich18,Abdujabbarov11,Okabayashi20}.

In this paper we shall consider external magnetic fields, due to the fact that black holes can not have their own magnetic fields~\cite{Ginzburg1964,Anderson70}. 
When thinking about astrophysical black holes, the external magnetic field may be due to the accretion disc around the black hole itself~\cite{Wald74} or to the existence of nearby neutron stars~\cite{Ginzburg1964,Rezzolla01,deFelice03,deFelice04}. 
The analysis of magnetic field strengths induced by different sources has been done in~\cite{Piotrovich10,Eatough13,Shannon13,Dallilar2018,Baczko16}. 
In the paper we also consider the simple case of a test magnetic field which does not modify the background geometry. 

In some of the galactic microquasars (binary systems containing a black hole surrounded by an accretion disk), twin high-frequency quasiperiodic oscillations (HF QPOs) are observed, usually, in the frequency ratio 3:2 indicating the presence of resonant phenomena \cite{Kluzniak01}.
There is a large variety of HF QPO models (see e.g. \cite{Stuchlik13A&A}) related to the epicyclic motion of hot spots of the accretions disks e.g. relativistic precession model \cite{Stella99-qpo}
or oscillatory models of the accretion disk \cite{Rezzolla_qpo_03a,Torok05A&A}.
In these models, the observed frequencies are related to the frequencies of the geodesic epicyclic motion, or to their combinations. However, it has been demonstrated that there is no unique model of this kind that could explain the twin 3:2 HF QPOs observed in three microquasars~\cite{Torok11A&A}. 
For this reason, the frequencies of the epicyclic motion of charged particles orbiting a magnetized black hole were considered \cite{Tursunov20ApJ,Panis19}
and it has been shown that the magnetic modifications of the epicyclic oscillation models are able to explain data observed in all three microquasars \cite{Kolos17-qpo,Stuchlik20}.
In the present article we perform a similar analysis for the case of $4D$ EGB black holes. We show that for three microquasars we can find suitable parameter values and thus explain the effects associated with epicyclic oscillation around $4D$ EGB black holes. 

The paper is organized as follows: In Sec.~\ref{Sec:metic} we briefly describe $4D$ EGB gravity and its black hole solutions. In Secs.~\ref{Sec:motion} and \ref{Sec:ep-freq} we study the charged particle and epecyclic motions with QPOs in the black hole vicinity in the presence of an external asymptotically uniform magnetic field. The effect of the Gauss-Bonnet coupling constant on the extracted energy by the collision process is studied in Sec.~\ref{Sec:energy}. We end up our concluding remarks in Sec.~\ref{Sec:conclusion}.

In this work we use a system of units in which $G=c=1$. Greek indices are taken to run from 0 to 3, Latin indices from 1 to 3.

\section{\label{Sec:metic}
	Black holes in $4D$ Einstein-Gauss-Bonnet gravity }

In $D=4$ dimensions, the action for the Gauss-Bonnet theory is given by
\begin{eqnarray}
S&=&\frac{1}{16\pi G}\int\sqrt{-g} d^4x\left[R+\alpha L_{GB}\right]\, ,
\end{eqnarray}
with $\alpha$ being the Gauss-Bonnet (GB) coupling constant and the GB contribution to the action given by 
\begin{eqnarray}
L_{GB}=R_{\mu\nu\lambda\delta}R^{\mu\nu\lambda
	\delta} - 4R_{\mu\nu}R^{\mu\nu}+R^2\, ,
\end{eqnarray}
where $R$ refers to the scalar curvature. 
Lovelock's theorem states that it is impossible for the GB term to contribute to the gravitational dynamics in $D=4$ since this term in the Lagrangian is a total derivative. However in \cite{Glavan20prl} it was shown that there exist a way to consider a non-trivial contribution to the equations of motion by rescaling the coupling constant thus avoiding Lovelock's theorem. As mentioned, the trick works at the level of equations of motion but does not provide a valid action for the 4-dimensional theory. In fact, by rescaling the coupling constant $\alpha\rightarrow \alpha/(D-4)$ in the Gauss-Bonnet term and then imposing the limit $D\rightarrow4$, the authors of \cite{Glavan20prl} were able to obtain spherically symmetric $4D$ EGB black hole solutions as  
\begin{eqnarray}\label{Eq:metric}
ds^2=-F(r)dt^2+\frac{dr^2}{F(r)}+r^2d\Omega^2,
\end{eqnarray}
with
\begin{eqnarray}\label{Eq:fr}
F(r)=1+\frac{r^2}{2\alpha}\left(1\pm\sqrt{1+\frac{8\alpha M}{r^3}}\right)\, ,
\end{eqnarray}
with black hole mass $M$.  
It is worth noticing that differently from GR in GB theory there exist two separate branches of black hole solutions depending on the sign in front of the square root. One solution, the one with the minus sign, has an attractive massive point source and asymptotically exhibits Schwarzschild-like behaviour
\cite{Dadhich07gb,Torii05}. 
This solution may mimic a Schwarzschild black hole for far away observer and therefore in the following we will focus of the `minus' branch of solutions.

In the limit $\alpha \rightarrow 0$, Eq.~(\ref{Eq:fr}) takes the form   
\begin{equation}
\lim_{\alpha\rightarrow 0} F(r) = 1-\frac{2M}{r}+\frac{4M^2}{r^4}\alpha+... 
\end{equation}
and it is easy to see that one can retrieve the Schwarzschild case when $\alpha \rightarrow 0$. Therefore this appears to be the right branch to investigate the departure from GR solution in the $4D$ EGB theory. 
From the first derivative of $F(r)$ we find one extremum point when $\alpha \neq 0$, i.e. $r=M^{1/3} \alpha^{1/3}$ given by
\begin{eqnarray}
F(r)=1-M^{2/3}\alpha^{-1/3}\, . 
\end{eqnarray}
This helps one to find the allowed range of the coupling constant $\alpha$. 
We then get $0<\alpha/M^2\leq 1$ by imposing the minimum condition, i.e. $F(\alpha^{1/3})\leq 0$. Therefore we will deal with  $0<\alpha/M^2\leq 1$ throughout the paper. 
However, it is worth noting that one can also consider $\alpha<0$.
It has been shown that values of the coupling constant $\alpha/M^2$ in the range $(-8,1)$ allow for the existence of a black hole and possible pathologies are hidden below the horizon (see for example \cite{Guo20arXiv}). 
The study of the stability of the solution further shows that the object is unstable for \mbox{$0<\alpha /M^2\leq 0.15$} and stable for \mbox{$-2 < \alpha/M^2 <0$} while stability/instability is not determined for \mbox{$\alpha/M^2 \leq -2$} (see for example \cite{Konoplya20arXiv}). 
However, for negative values of $\alpha$ the connection of the EGB theory with Einstein's theory is not straightforward. For this reason in the following we focus on the case of $\alpha>0$.  
Let us then consider a coordinate singularity for the black hole in $4D$ EGB gravity. By imposing the condition $F(r)=0$ we determine the horizon of the black hole as 
\begin{eqnarray}\label{Eq:hor}
r_{h}=M \pm \sqrt{M^2-\alpha }\, .
\end{eqnarray}
It is immediately clear that the two outer and inner horizons coincide if and only if  $\alpha=M^{2}$ and an extremal black hole with horizon $r=M$ is obtained in this case. Note that the black hole horizon no longer exists in the case $\alpha>M^{2}$, in which case the space-time exhibits a naked singularity.
{ Also, the presence of the GB term has the physical effect of shifting the outer horizon, as well as the innermost stable circular orbit, inwards, closer to the central singularity. This is consistent with the interpretation of the GB term as a repulsive gravitational charge, that physically manifests in a change of the causal structure similar to that of the introduction of the electric charge in classical relativistic black holes (see for example \cite{Dadhich20egb}).}

\section{\label{Sec:motion}
	Charged particle motion }

Here we consider charged particle motion in the gravitational field of the $4D$ EGB black hole immersed in an external asymptotically uniform magnetic field.  The existence of a timelike $\xi^{\mu}_{(t)}=(\partial/\partial
t)^{\mu}$ and a spacelike $\xi^{\mu}_{(\varphi)}=(\partial/\partial \phi)^{\mu}$ Killing vector 
allows one to take two Killing equations as~\cite{Wald74}  
\begin{eqnarray}\label{Eq:Killing}
\label{kelling} \xi_{\mu; \nu}+\xi_{\nu; \mu}= 0\, ,
\end{eqnarray}
so that we have $ \Box \xi^\mu=0$ in the vacuum case, similarly 
to the case of Maxwell equations in the Lorentz
gauge for which $ \Box A^\mu=0$, where $A^\mu$ is the electromagnetic 4-potential. However, the line element of $4D$ EGB gravity spacetime is not Ricci flat, i.e. $ \Box \xi^\mu=\xi^{\mu;\nu}_{\ \ \ ;\nu}=R^{\mu}_{\delta}\xi^{\delta}\neq0$. Thus, the vector potential of the electromagnetic field must take the form 
\begin{eqnarray}\label{Vec-pot}
A^\mu=C_1 \xi^\mu_{(t)} +C_2 \xi^\mu_{(\phi)}+n^{\mu}\, ,
\end{eqnarray}
where integration constants can be considered as $C_1=0$ and $C_2=B/2$, respectively, for a static and spherically symmetric black hole spacetime and the vector $n^{\mu}$ accounts for the non-flat Ricci tensor $R_{\mu\nu}\neq0$. Then $n^\mu$ and $R_{\mu\nu}$ are related by the following expression 
\begin{eqnarray}
\Box n^{\mu}=C_{2} \xi^{\gamma}_{(\phi)}R^{\mu}_{\gamma}\, .
\end{eqnarray}
Hence, we obtain the components of the electromagnetic field's 4-vector potentials as
\begin{eqnarray}
\label{4pot} A_t&=&A_{r}= A_{\theta}=0\, , ~~\mbox{and}~~ A_{\varphi}=\frac{B}{2}r^2\left(1+\frac{3M^2}{5r^4}\alpha\right)\sin^2\theta 
\, . 
\end{eqnarray}
\begin{figure*}
	\centering
	\includegraphics[width=0.45\textwidth]{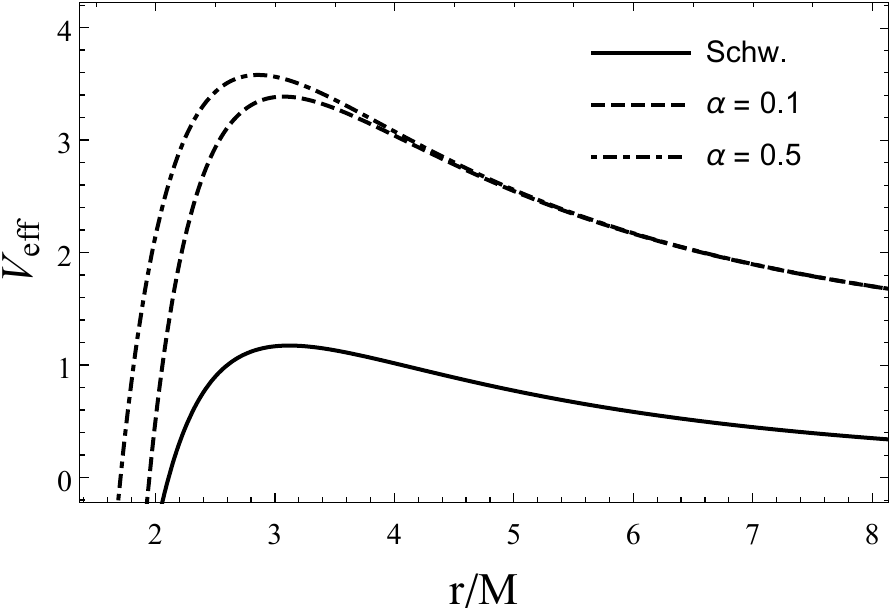}
	\includegraphics[width=0.45\textwidth]{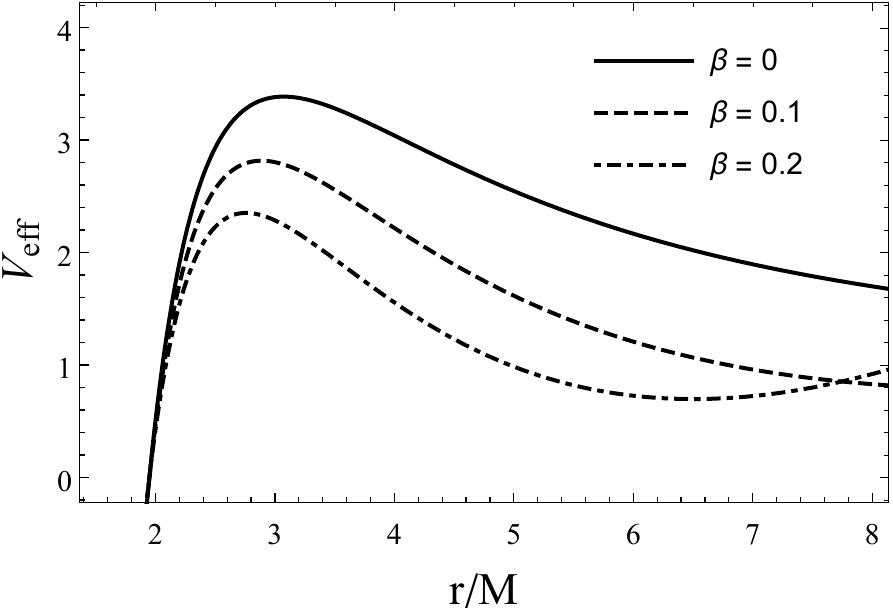}
	
	\caption{\label{fig1} Radial dependence of the effective potential for massive particles around a black hole in $4D$ EGB gravity and immersed in an external asymptotically uniform magnetic field. From left: $V_{eff}$ is plotted for different values of $\alpha$ in the case of $\beta=0$ and for different values of $\beta$ in the case of $\alpha=0.1$. }
\end{figure*}
Given the four velocity of zero angular momentum observers (ZAMO) $(u_{\mu})_{_{\textrm{ZAMO}}}=
\left\{-\sqrt{F},0,0,0\right\}$  the Faraday tensor components are written as 
\begin{eqnarray}\label{Eq:faraday1}
\mathcal{F}_{r\phi}&=& B r \left(1-\frac{3M^2}{5r^4}\alpha\right)\sin^2\theta\, ,\\
\mathcal{F}_{\theta
	\phi}&=&B r^2 \left(1+\frac{3M^2}{5r^4}\alpha\right)\sin\theta  \cos\theta\, .
\label{Eq:faraday2}
\end{eqnarray}
It is then straightforward to determine the physical components of the magnetic field obtained by projecting the magnetic field components on the tetrads evaluated by ZAMO. These components are 
\begin{eqnarray}
\label{M1}  
B^{\hat r}=-B\left(1+\frac{3M^2}{5r^4}\alpha\right)\cos\theta \,  ~~\mbox{and}~~ 
B^{\hat\theta} =B\sqrt{F}\left(1-\frac{3M^2}{5r^4}\alpha\right)\sin\theta \, .
\end{eqnarray}

In flat spacetime and at asymptotic infinity, ie. for $M/r\rightarrow 0$ Eq.~
(\ref{M1}) yields  
\begin{eqnarray}
\label{Eq:limit} && B^{\hat r} =-B\cos\theta\, , ~~ B^{\hat\theta}=B\sin\theta\, , 
\end{eqnarray}
which describes a homogeneous magnetic field in flat spacetime.

To study of the motion of charged particles in the vicinity of the $4D$ EGB black hole and in the presence of an external magnetic field is done by considering the Hamiltonian-Jacobi equation~\cite{Misner73} 
\begin{eqnarray}
H  \equiv \frac{1}{2}g^{\mu\nu}\left(\frac{\partial S}{\partial x^{\mu}} - q A_\nu\right)\left(\frac{\partial S}{\partial x^{\nu}} - q A_\nu\right)\, ,
\label{Eq:H}
\end{eqnarray}
with the action $S$, the coordinate four-vector $x^\mu$,
the electromagnetic field four-vector potential $A_\alpha$ from Eq.~(\ref{4pot}) and the test particle's charge $q$. The Hamiltonian is a constant that can be set to $H=k/2$ with $k=-m^2$ (where $m$ is the test particle's mass).  

Then from Hamilton-Jacobi equation, using the conservation equations for the two Killing vectors, we obtain the action $S$ for the motion of charged particles around the black hole as 
\begin{eqnarray}\label{Eq:action}
S= \frac{1}{2}k\lambda-Et+L\varphi+S_{r}(r)+S_{\theta}(\theta)\ .
\end{eqnarray}
Here $E $ and $L$ are the energy and angular momentum of the charged particle,  respectively. From Eq. (\ref{Eq:action}), we rewrite Hamilton-Jacobi equation in the following form 
\begin{eqnarray}\label{Eq:separable}
&k=& -F(r)^{-1}E^2
+ F(r)
\left(\frac{\partial S_{r}}{\partial r}\right)^2
+\frac{1}{r^2}\left(\frac{\partial S_{\theta}}{\partial \theta}\right)^2+
\frac{\left(L-qA_{\varphi}\right)^2}{r^2\sin^2\theta} \, , 
\end{eqnarray}
\textcolor{black}{
	and introduce two Hamiltonian parts, dynamical and potential as
	\begin{eqnarray}
	H_{\textrm{dyn}}&=& \frac{1}{2}\left(F(r)
	\left(\frac{\partial S_{r}}{\partial r}\right)^2 +\frac{1}{r^2}\left(\frac{\partial S_{\theta}}{\partial \theta}\right)^2\right),\\
	H_{\textrm{pot}}&=& \frac{1}{2}\left(-F(r)^{-1}E^2+\frac{\left(L-qA_{\varphi}\right)^2}{r^2\sin^2\theta}-k\right).\label{Eq:sepham}
	\end{eqnarray}}
The system has four independent constants of motion, three of which ($E$, $L$ and $k$) have been specified. The fourth constant of motion can be obtained due to separability of the action and it is related to the latitudinal motion of test particles
\cite{Misner73}. 
However, in the following we will restrict the attention to equatorial motion, setting $\theta=\pi/2$ and therefore we can ignore the fourth constant of motion.
\begin{table*}
\caption{\label{tab1} The values of the ISCO radius $r_{isco}$ are tabulated in the case of charged particles moving around $4D$ EGB black hole for different values of GB coupling constant $\alpha$ and magnetic parameter $\beta$. }
\begin{tabular}{c|cccccc}
$$ & $$ & $$ &  $$
& $\beta$ & $$ & $$ \\ \hline
{$\rm \alpha $} & $0.000$ & $0.001$ &  $0.005$
& $0.010$ & $0.050$ & $0.100$ \\
    & $$ & -0.001  &  -0.005 &  -0.010 &  -0.050 &  -0.100 \\
\hline
 0.00 & 6.00000 &5.99826 & 5.95709 & 5.84140 & 4.69667 & 3.98268 \\
    & $$         &5.99829 &5.95986 &5.85955 &5.02571 &4.63195 \\\\
0.01  &5.99388  &5.99215  &5.95109  &5.83571 &4.69245 &3.97865 \\
      & $$      &5.99217  &5.95385  &5.85381  &5.02112 &4.62743 \\\\
0.05  &5.96918  &5.96747 &5.92691 &5.81276 &4.67538 &3.96239 \\
& $$            &5.96749 &5.92963 &5.83066 &5.00259 &4.60921 \\\\
0.1 &5.93782   &5.93614 &5.89619 &5.78359 &4.65365 &3.94167 \\
      & $$     &5.93616 &5.89887 &5.80124 &4.97900 &4.58602 \\ \\
0.5 &5.66395   &5.66248 &5.62750 &5.52762 &4.46070 &3.75646 \\
      & $$     &5.66250 &5.62983 &5.54320 &4.77049 &4.38067 \\\\
1.0 &5.23655   &5.23535 &5.20681 &5.12376 &4.14430 &3.44333 \\
      & $$     &5.23537 &5.20869 &5.13662 &4.43335 &4.04561 \\
\end{tabular}
\end{table*}
\begin{figure}
	\centering
	\includegraphics[width=0.4\textwidth]{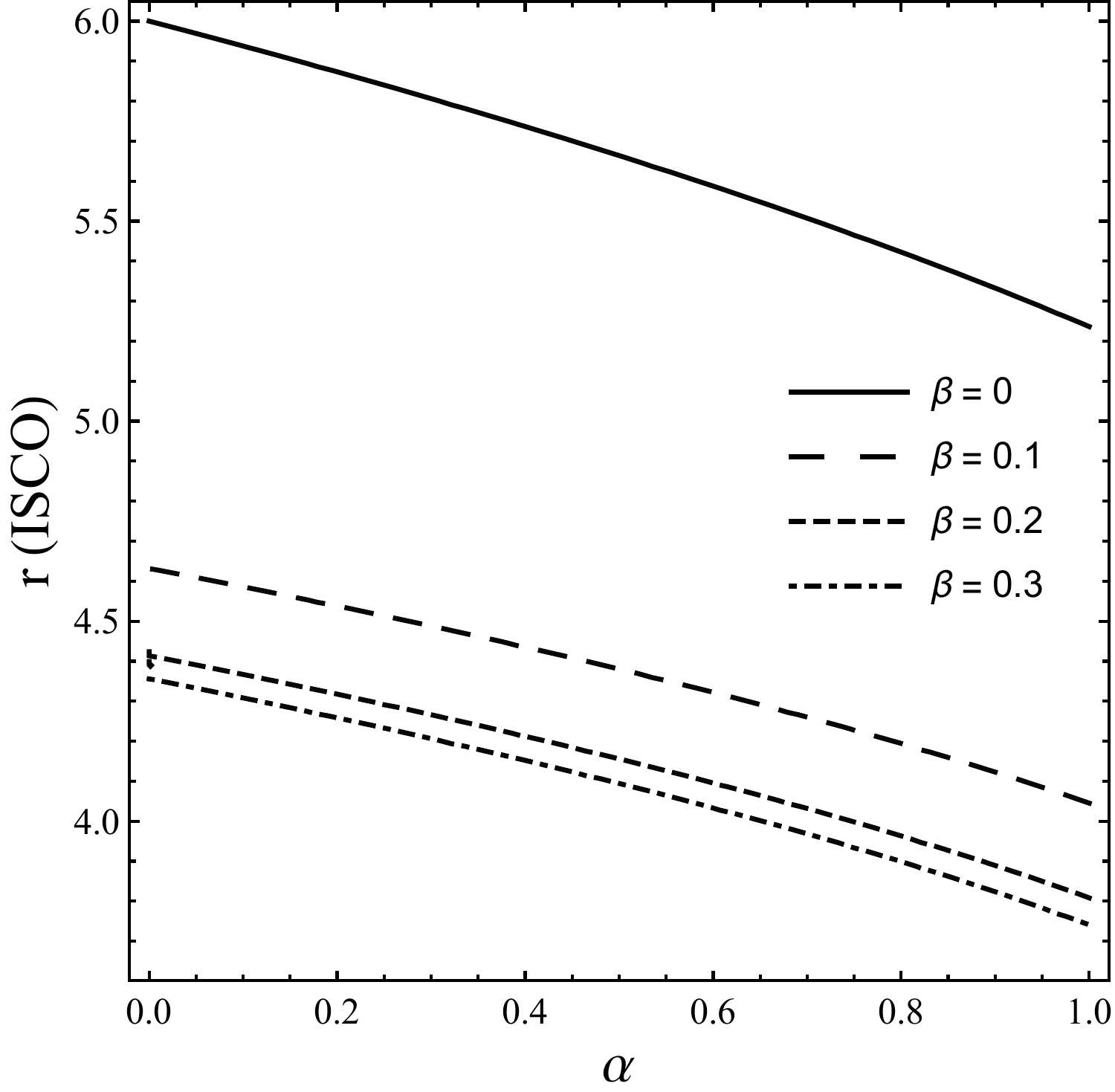}%
	\caption{\label{fig2} The dependence of the ISCO radius on the GB coupling constant $\alpha$ for massive particle around $4D$ EGB black hole immersed in an external uniform magnetic field. The ISCO radius is plotted for different values of $\beta$.
	}
\end{figure}
From Eq.~(\ref{Eq:separable}) we obtain the radial equation motion for charged particles as 
\begin{eqnarray}
\frac{1}{2}\dot{r}^{2} + V_{eff}(r;\mathcal{L},\alpha,\beta)=\mathcal{E}^2\, ,
\end{eqnarray}
where the dot denotes derivative with respect to the proper time of the particle $\tau$ and the radial function $V_{eff}(r;\mathcal{L},\alpha,\beta)$ is the effective potential of the system which is given by 
\begin{eqnarray} \label{Eq:Veff}
V_{eff}(r;\mathcal{L},\alpha,\beta) &=&  \left[1+\frac{r^2}{2\alpha}\left(1-\sqrt{1+\frac{8\alpha M}{r^3}}\right)\right]\nonumber\\&\times&\left\{
1+\frac{1}{r^2}\left[\mathcal{L}-\frac{\beta}{M}~\left(r^{2}+\frac{3M^2}{5r^2}\alpha\right)\right]^2\right\}\, , 
\end{eqnarray}
with the conserved constants per unit mass given by $\mathcal{E}=E/m$ and
$\mathcal{L}=L/m$ and $k/m^2=-1$. The magnetic field parameter, reintroducing $G$ and $c$ for convenience in evaluating physical scenarios, is defined as
\begin{eqnarray}\label{beta}
\beta\equiv\frac{qBMG}{mc^4}\, ,
\end{eqnarray}
and it measures the influence of the external magnetic field on the motion of charged particles. It is easy to see that in the case of small values of $\alpha$ and $\beta$ the effective potential tends to the Schwarzschild case.

The radial dependence of the effective potential (\ref{Eq:Veff}) for different values of $\alpha$ and $\beta$ is shown in Fig.~\ref{fig1} from which it can be seen that the GB term and the magnetic field have opposite effects.

Next, let us come to study circular orbits of charged particles around a black hole in $4D$ EGB gravity in the presence of external asymptotically uniform magnetic field. By imposing the following conditions for the effective potential and its first derivative  
\begin{eqnarray}\label{Eq:en}
\mathcal{E}^2=V_{eff}(r;\mathcal{L},\alpha,\beta)\, , \\ \nonumber\\
\label{Eq:an}
\frac{\partial V_{eff}(r;\mathcal{L},\alpha,\beta)}{\partial r}=0\, .
\end{eqnarray}
we obtain the values of the constants of motion $\mathcal{E}$ and $\mathcal{L}$ at circular orbits. Thus, the angular momentum for particles on circular orbit is  given by 
\begin{eqnarray}\label{Eq:L}
\mathcal{L}&=& \frac{1}{r F'(r)-2 F(r)}\Big\{\beta r^3\left[F'(r)N(r)+F(r)N'(r)\right] +
\nonumber\\&+&\left. r^{3/2}\Big[\beta^2 r F(r)^2 \big(r N'(r)+2 N(r)\big)^2-r F'(r)^2+2 F(r) F'(r)\Big]^{1/2}\right\}\, , 
\end{eqnarray}
where primed quantities denote derivatives with respect to $r$ and $N(r)=1+{3M^2\alpha}/{5r^4}$. The innermost stable circular orbit (ISCO) is then determined from the condition 
\begin{eqnarray}\label{Eq:ISCO}
\frac{\partial^2 V_{eff}(r;\mathcal{L},\alpha,\beta)}{\partial r^2}=0\, .
\end{eqnarray}
In Fig.~\ref{fig2} we show the dependence of ISCO radius on the GB coupling $\alpha $ for the different values of $\beta$. We see that the ISCO radius in $4D$ EGB gravity becomes smaller as the effect of the GB term increases.
{This is due to the already mentioned fact that the GB term acts as a repulsive gravitational charge, thus weakening the strength of the gravitational field at a fixed distance, in turn allowing for circular orbits to remain stable closer to the source.}
Note that the ISCO radius also decreases as a consequence of the introduction of a magnetic field. 
In table~\ref{tab1}, we provide numerical values for the ISCO radius for the different values of the GB coupling constant and magnetic field parameter. 

\section{\label{Sec:energy}
	Energetic collisions }

We consider now the collision energy of two particles in the geometry of the $4D$ EGB black hole. 
We assume that the two particles have rest masses
$m_{1}$ and $m_2$ at spatial infinity.  
The four-momentum and the total momenta of the two colliding particles
$(i =1, 2)$ are given by
\begin{eqnarray}\label{Eq:4-mom1}
p_{i}^{\alpha}&=&m_{i}u_{i}^{\alpha}, \\
\label{Eq:4-mom2}
p_{t}^{\alpha}&=&p_{1}^{\alpha}+p_{2}^{\alpha}\,.
\end{eqnarray}
where $u_{i}^{\alpha}$ is the four velocity of the particle $i$. From the Hamilton's equation of motion the four-momentum is given by   
\begin{eqnarray}\label{Eq:4-mom3}
p^{\alpha}=g^{\alpha\beta}\left(\pi_{\beta}-qA_{\beta}\right)\, ,
\end{eqnarray}
where $\pi_{\beta}$ is the canonical four-momentum of a charged particle.
Based on the Eqs.~(\ref{Eq:4-mom1}) and (\ref{Eq:4-mom2}) the center of mass energy $E_{cm}$ of the collision between the two
particles is given by \cite{banados09}
\begin{eqnarray}\label{Eq:cm1}
\frac{E_{cm}^{2}}{2m_1 m_2}=\frac{m_{1}^2+m_{2}^2}{2m_1
	m_2}-g_{\alpha\beta}u^{\alpha}_{1}u^{\beta}_{2}\,.
\end{eqnarray}
Our aim is to understand the impact of the GB coupling constant $\alpha$ on the energy extracted from the collision process and compare to Einstein gravity. In the context of astrophysics the mechanism could help explain the observed production of high energy particles by black hole candidates. 
For simplicity we consider particle collisions 
of two free falling particles which occurs near the horizon of the black hole. Employing equation (\ref{Eq:4-mom3}) for the particle $i$ 
and substituting them into the general form of center of mass energy (\ref{Eq:cm1}), we get 
\begin{eqnarray}
\label{Eq:cm2} \frac{E^{2}_{\rm cm}}{2m_{1}m_{2}} 
& = &1+\frac{\left(m_{1}-m_{2}\right)^2}{2m_1
	m_2}+\frac{\mathcal{E}_1 \mathcal{E}_2}{F(\tilde{r})}-\nonumber\\&-&
	\frac{\left(\tilde{L}_1-\beta \left(\tilde{r}^2+\frac{3\tilde{\alpha}}{5\tilde{r}^2}\right)\right) \left(\tilde{L}_2-\beta \left(\tilde{r}^2+\frac{3\tilde{\alpha}}{5\tilde{r}^2}\right)\right)}{\tilde{r}^2}-\nonumber\\
&-&\frac{1}{F(\tilde{r})}\sqrt{\mathcal{E}_1^2-F(\tilde{r}) \left[\frac{\left(\tilde{L}_1-\beta \left(\tilde{r}^2+\frac{3\tilde{\alpha}}{5\tilde{r}^2}\right)\right)^2}{\tilde{r}^2}+1\right]}\nonumber\\
&\times &  \sqrt{\mathcal{E}_2^2-F(\tilde{r}) \left[\frac{\left(\tilde{L}_2-\beta \left(\tilde{r}^2+\frac{3\tilde{\alpha}}{5\tilde{r}^2}\right)\right)^2}{\tilde{r}^2}+1\right]}\, ,
\end{eqnarray}
where we have defined $\tilde{L}_{1,2}=\mathcal{L}_{1,2}/M$, $\tilde{\alpha}=\alpha/M^2$ and $\tilde{r}=r/M$. 
Eq. (\ref{Eq:cm2}) describes the center of mass energy for collision of two charged particles having respectively two different specific angular momenta $\tilde{L}_{1,2}$ and specific energies $\mathcal{E}_{1,2}$.  

\begin{figure}
	\centering
	\includegraphics[width=0.45\textwidth]{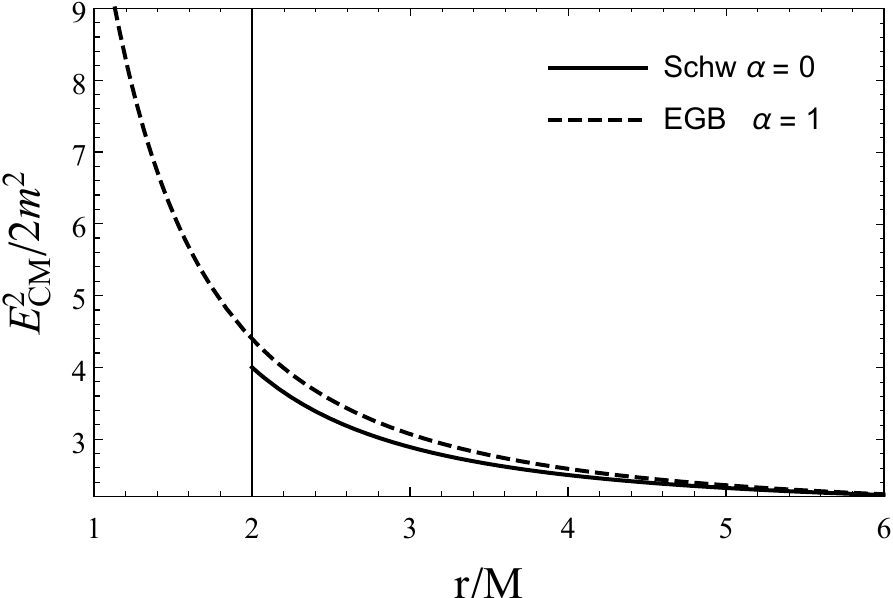}
	
	\caption{\label{fig3} The center of mass energy $E_{\rm cm}$ for the collision of two particles moving around a black hole with angular momenta $\tilde{L}_{1}$ and $\tilde{L}_{2}$, respectively, in the absence of magnetic field (i.e. $\beta=0$). The solid line corresponds to the Schwarzschild black hole in Einstein's gravity (the event horizon is located at $r/M=2$ in this case), while the dashed line corresponds to the 4D EGB black hole with $\alpha=1$ (the event horizon is located at $r/M=1$ in this case). For this figure, we used the particular case in which $\mathcal{E}_1=\mathcal{E}_2=1$ and $\tilde{L}_1=-\tilde{L}_2=2$. Note that $E_{\rm cm}$ is larger for a black hole in EGB gravity in comparison to the Schwarzschild case. However, one can see that in both cases $E_{\rm cm}$ is finite at the horizon.}
\end{figure}

From Fig.~\ref{fig3} it can be seen that the center of mass energy increases as the distance from the source decreases. Therefore, we consider the limiting case $\tilde{r}\rightarrow \tilde{r}_{h}$ for which the center of mass energy becomes the highest. For the sake of clarity, we further assume $m_1=m_2=m$ for free falling particles and evaluate the limiting value for the center of mass energy extracted by particle collision at the 
horizon
\begin{eqnarray}\label{Eq:cm3}
\frac{E^{2}_{\rm cm}(\tilde{r}\rightarrow \tilde{r}_{h})}{m^2}
&=&2+2 \left(\tilde{L}_1+\tilde{L}_2\right)\beta \nonumber\\&+&\frac{\mathcal{E}_2 \left[1+(\beta-\tilde{L}_1) \left(2 \sqrt{1-\tilde{\alpha} } (\beta+\tilde{L}_1)+\beta-\tilde{L}_1\right)\right]}{\mathcal{E}_1}\nonumber\\&+&\frac{\mathcal{E}_1 \left[1+(\beta-\tilde{L}_2) \left(2 \sqrt{1-\tilde{\alpha} } (\beta+\tilde{L}_2)+\beta-\tilde{L}_2\right)\right]}{\mathcal{E}_2}\nonumber\\&-& 2 \left(1-2 \sqrt{1-\tilde{\alpha} }\right) \tilde{L}_1 \tilde{L}_2-2 \left(1+2 \sqrt{1-\tilde{\alpha} }\right) \beta^2\, .
\end{eqnarray}
Now, considering $\mathcal{E}_1=\mathcal{E}_2$, from Eq.~(\ref{Eq:cm3}) we get the simple formula 
\begin{eqnarray}
\label{Eq.cm4} \frac{E^{2}_{\rm cm}(\tilde{r}\rightarrow \tilde{r}_{h})}{m^2}
&=& \Big[4+\left(1- 2\sqrt{1-\tilde{\alpha} }\right)\left(\tilde{L}_1-\tilde{L}_2\right)^2\Big] \, .
\end{eqnarray}
From the above equation it is immediately clear that the effect of magnetic field on the extracted energy does not participate in the limiting case when $\mathcal{E}_1=\mathcal{E}_2$.  
Also it is easy to notice that the center of mass energy extracted by two particle collisions at the horizon increases with increasing the coupling parameter $\alpha$ as the horizon gets closer to the source. The behaviour of $E_{\rm cm}$ as a function of $r$ is also shown in Fig.~\ref{fig3}. It is clear that  the center of mass energy for  black hole in $4D$ EGB gravity remains finite at the horizon and it is larger for $\alpha>0$ as compered to the one for Schwarzschild case. 

\section{\label{Sec:ep-freq}
	Epicyclic frequencies and mass-limit of microquasars}

We consider here a charged particle with stable circular motion, i.e. motion at the minimum of the effective potential which is only allowed for $r\geq r_{isco}$. If such a particle is slightly displaced from the circular radius $r_0$, the so-called epicyclic motion occurs around the equilibrium position. For a small displacement $r = r_0 + \delta r$ and $\theta=\pi/2+\delta \theta$, 
the particle oscillates just like a linear harmonic oscillator and the displacements are governed by the equations
\begin{eqnarray}
\label{harmosc}
\delta\ddot{ r}+\bar{\omega}_r^2 \delta r = 0, \ \ \ 
\delta\ddot{ \theta}+\bar{\omega}_\theta^2 \delta\theta = 0,
\end{eqnarray}
where $\bar{\omega}_r$ and $\bar{\omega}_\theta$ denote the frequencies of the radial and latitudinal epicyclic oscillation respectively, as measured by a local observer. The orbital frequency $\bar{\omega}_\phi$ comes from the definition of angular momentum. These frequencies are then determined by equations
\begin{eqnarray}
\label{baromegas}
\bar{\omega}_r^2 &=& \frac{1}{g_{rr}}\frac{\partial^2 H_{\textrm{pot}}}{\partial r^2},\\ \nonumber
\bar{\omega}_\theta^2 &=& \frac{1}{g_{\theta\theta}}\frac{\partial^2 H_{\textrm{pot}}}{\partial \theta^2},\\ \nonumber
\bar{\omega}_\phi &=& \frac{1}{g_{\phi\phi}}\Big(\mathcal{L}-q A_\phi\Big).
\end{eqnarray}

\begin{figure*}
	\centering
	\includegraphics[width=0.8\textwidth]{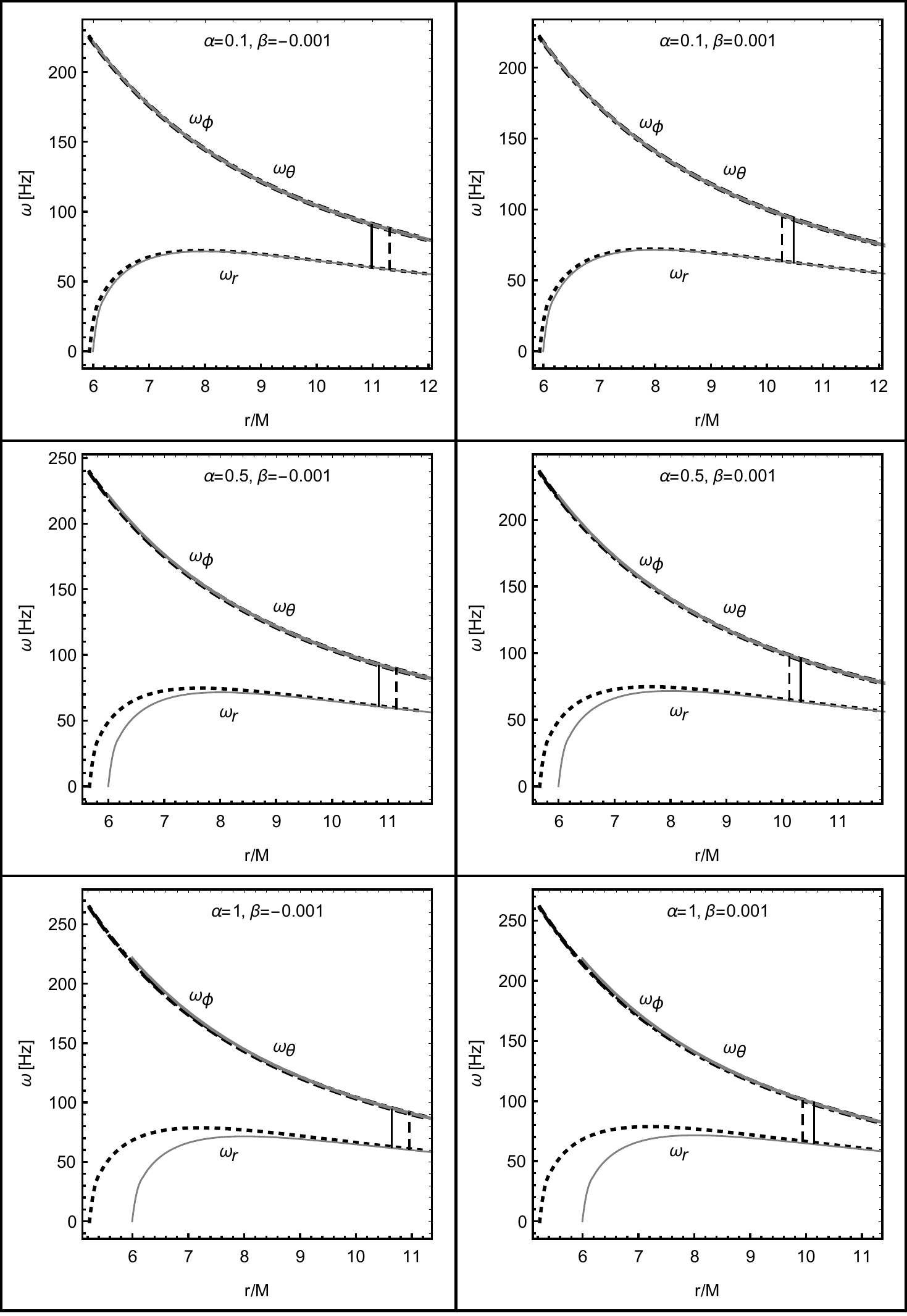}%
	\caption{\label{fig:ef0001} The epicyclic frequencies for various parameter $\alpha$ of $4D$ EGB black hole immersed in an external uniform magnetic field described by parameter $\beta$. Dotted curves are radial frequencies, dashed curves are latitudinal frequencies and dot-dashed are orbital frequencies. Thin gray lines denotes epicyclic frequencies for the Schwarzschild black hole immersed in an external uniform magnetic field. Vertical lines indicates a position of 3:2 resonance in between radial-latitudinal frequencies (solid), latitudinal-orbital frequencies (dotted) and radial-orbital (dashed). }
\end{figure*}

\begin{figure*}
	\centering
	\includegraphics[width=0.8\textwidth]{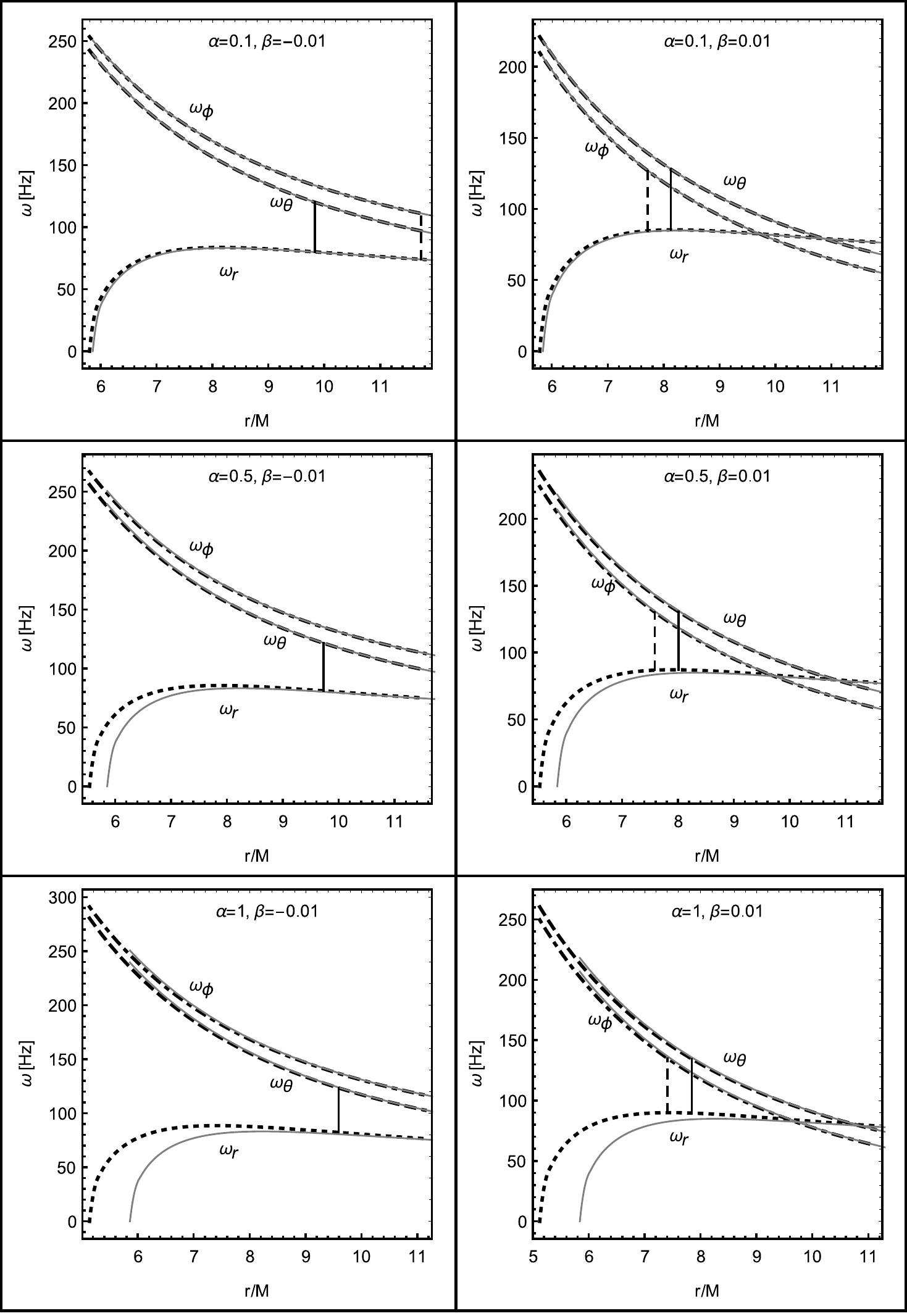}%
	\caption{\label{fig:ef001} The epicyclic frequencies for various parameter $\alpha$ of $4D$ EGB black hole immersed in an external uniform magnetic field described by parameter $\beta$. Dotted curves are radial frequencies, dashed curves are latitudinal frequencies and dot-dashed are orbital frequencies. Thin gray lines denotes epicyclic frequencies for the Schwarzschild black hole immersed in an external uniform magnetic field. Vertical lines indicates a position of 3:2 resonance in between radial-latitudinal frequencies (solid), latitudinal-orbital frequencies (dotted) and radial-orbital (dashed). }
\end{figure*}

\begin{figure*}
	\centering
	\includegraphics[width=0.8\hsize]{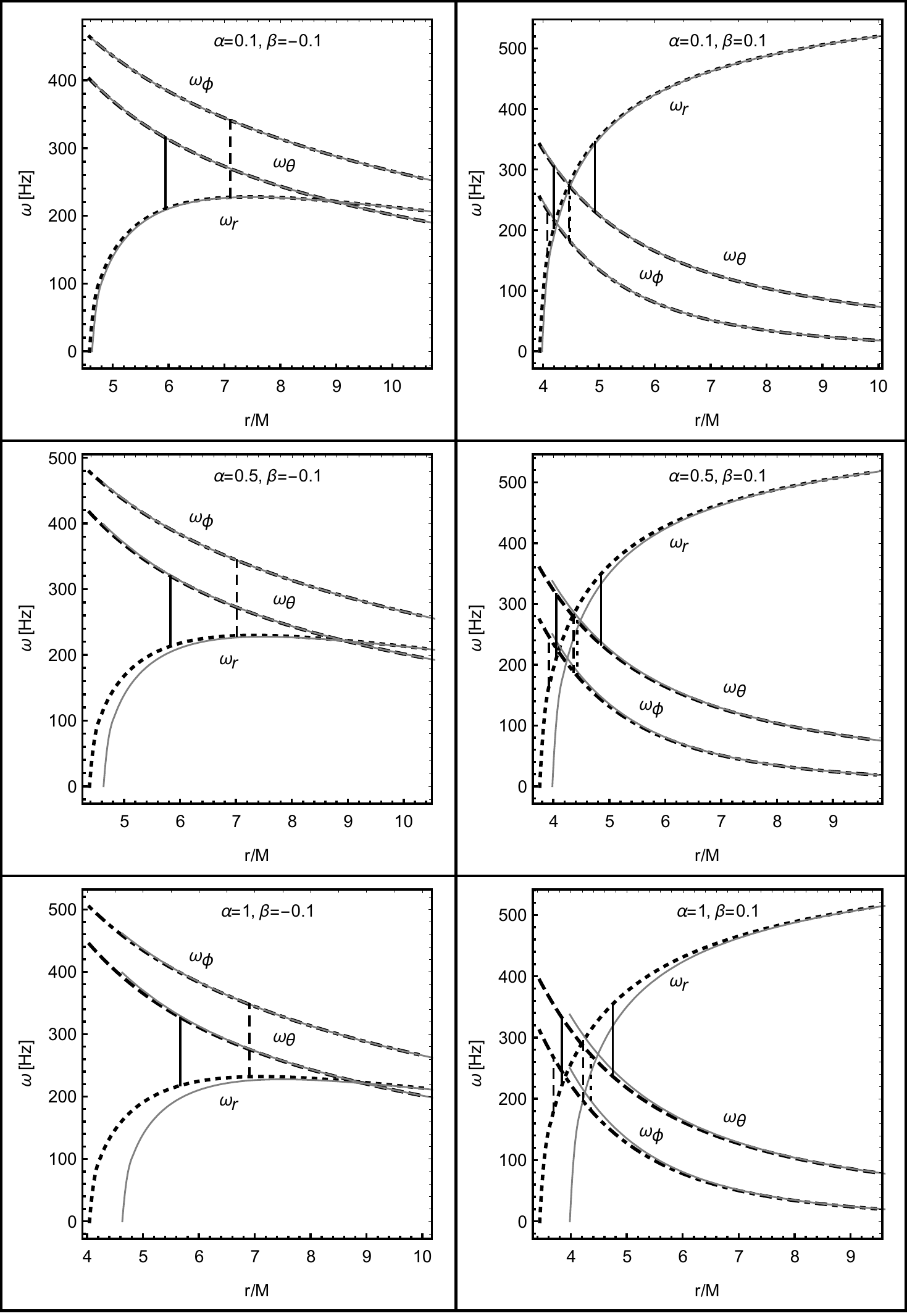}%
	\caption{\label{fig:ef01} The epicyclic frequencies for various parameter $\alpha$ of $4D$ EGB black hole immersed in an external uniform magnetic field described by parameter $\beta$. Dotted curves are radial frequencies, dashed curves are latitudinal frequencies and dot-dashed are orbital frequencies. Thin gray lines denotes epicyclic frequencies for the Schwarzschild black hole immersed in an external uniform magnetic field. Vertical lines indicates a position of 3:2 resonance in between radial-latitudinal frequencies (solid), latitudinal-orbital frequencies (dotted) and radial-orbital (dashed). }
\end{figure*}

For distant observers, these frequencies need to be transformed into the form that the observer measures at infinity. Locally measured frequencies ($\bar{\omega}$) and frequencies measured at infinity ($\omega$) are related by the transformation from the proper time $\tau$ to the time measured at infinity $t$. This, reintroducing $G$ and $c$, can be written as
\begin{eqnarray}\label{transfreq}
\omega=\frac{1}{2\pi}\frac{c^3}{GM}\frac{\bar{\omega}}{(-g^{tt})\mathcal{E}},
\end{eqnarray}
where for convenience we have reintroduced $G$ and $c$. 
In Figs. \ref{fig:ef0001} - \ref{fig:ef01}, we plotted the epicyclic frequencies for the $4D$ Gauss-Bonnet black holes immersed in an external uniform magnetic field (depending on both parameters $\alpha$ and $\beta$) and compared them to the frequencies for the Schwarzschild black hole immersed in an external uniform magnetic field (depending only on $\beta$). 

Obviously, the substantial difference between the frequencies in these spacetimes is based on the different position of the ISCO as $\alpha$ increases. This in turn leads to a 3:2 resonance shift.

\subsection{Mass-limit of microquasars}

In order to asses the possible validity of the 4D EGB theory in astrophysics one needs to compare some observable predictions of the theory with the corresponding results in Einstein's gravity in the hope that they may be distinguished. To this purpose quasi-periodic oscillations (QPOs) offer a valuable tool to test theoretical models.
In particular an important application in present astrophysics can be found in the high-frequency quasi-periodic oscillations (HF QPO) observed from many low-mass X-ray binaries (LMXB) containing a black hole or a neutron star. Occasionally, these HF QPOs are observed in pairs (upper and lower frequencies). For many black hole LMXBs we find a ratio of these frequencies $f_U:F_L$ (upper and lower frequency) fixed equal to 3:2 \cite{Torok05A&A,Remillard06ApJ}, which states that the upper frequencies are very close to the orbital frequencies of the ISCO, we can conclude that these phenomena occur in close proximity to the central object, i.e. in a regime of very strong gravity.

This ratio of twin HF QPO frequency \mbox{($f_U:f_L$ = 3:2)} is observed, for example, for the sources \mbox{GRO 1655-40}, \mbox{XTE 1550-564} and \mbox{GRS 1915 + 105}. One straightforward interpretation is to identify the frequency $f_U$ with $\omega_\theta$ and the frequency $f_L$ with $\omega_r$. Of course, there are also other possibilities. As it can be seen for example in Figs.~\mbox{\ref{fig:ef0001} - \ref{fig:ef01}}, however, this identification corresponds to epicyclic resonance model \cite{Torok11A&A,Stuchlik13A&A}.

\begin{figure*}
	\centering
	\includegraphics[width=1\textwidth]{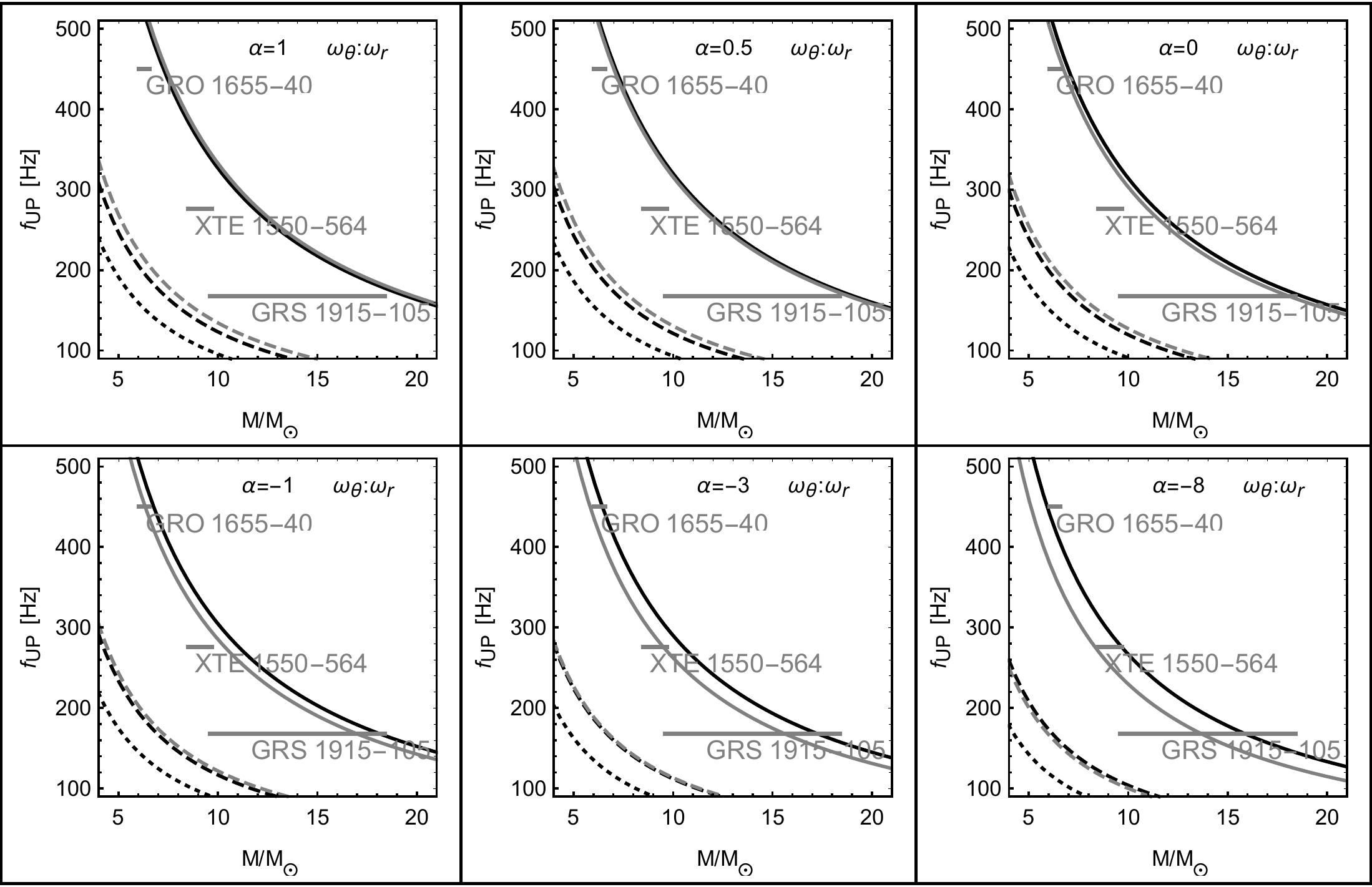}%
	\caption{\label{fig:massla} The upper frequency - mass relations (resonance 3:2 of $\omega_\theta:\omega_r$) for various parameter $\alpha$. The curves for parameter $\beta$ for positive values are \textit{black}, for negative values \textit{gray}. Dotted curves corresponds to $\beta=0$, dashed to $\beta=\{-0.01, 0.01\}$ and solid to $\beta=\{-0.1, 0.1\}$.}
\end{figure*}

\begin{figure*}
	\centering
	\includegraphics[width=1\textwidth]{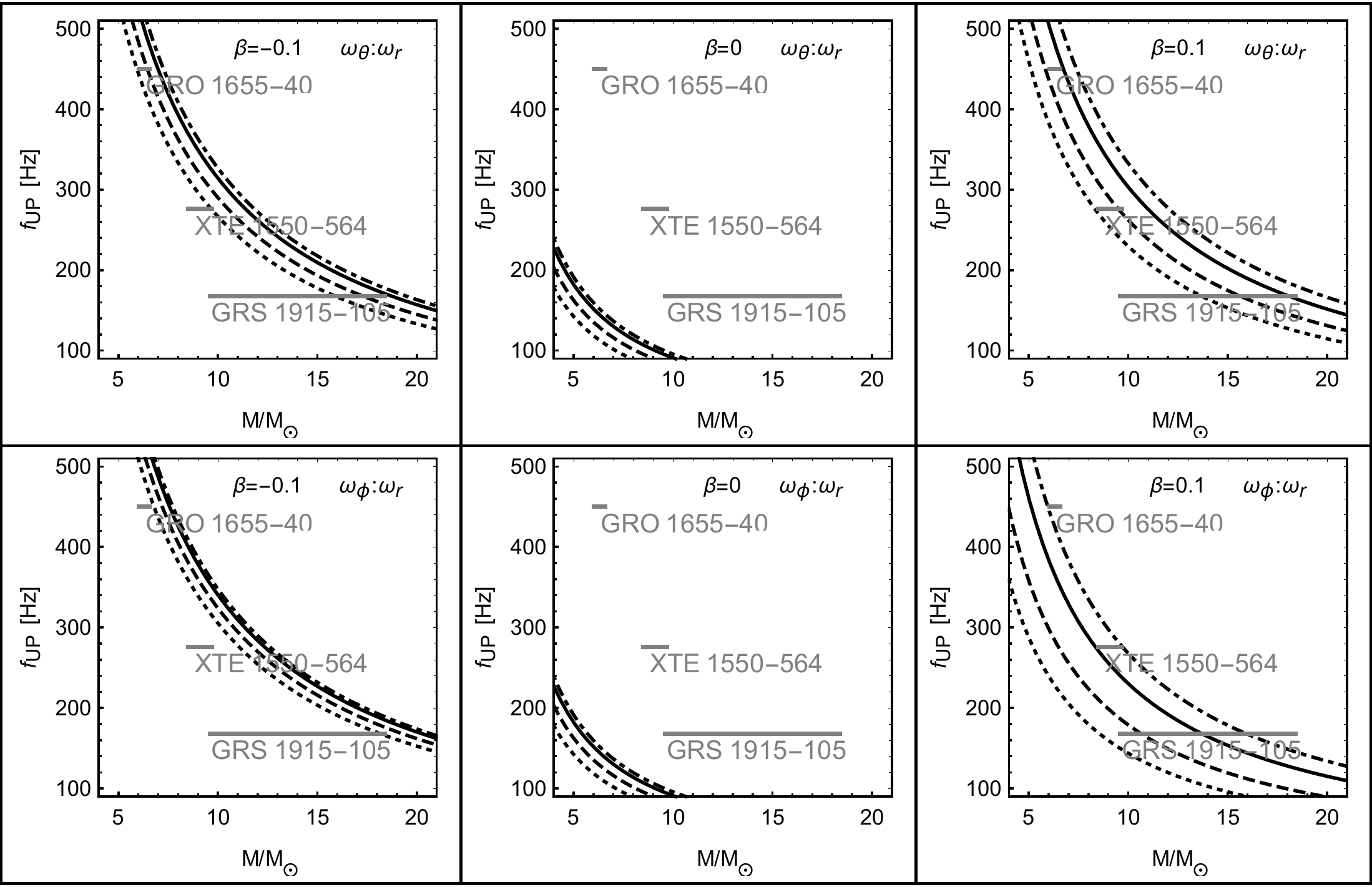}%
	\caption{\label{fig:masslb} The upper frequency - mass relations (resonance 3:2 of $\omega_\theta:\omega_r$ and $\omega_\phi:\omega_r$) for various parameter $\beta$. Solid curves corresponds to $\alpha=0$, dashed to $\alpha=-3$, dotted to $\alpha=-8$ and dotdashed to $\alpha=1$. }
\end{figure*}

The influence of the $\alpha$ parameter on the dependence of the upper frequency and mass with respect to astrophysical data is presented in Figs. \ref{fig:massla} and \ref{fig:masslb}. 
In order to obtain a good fit of the observed data, the presence of a non vanishing magnetic field is necessary. It can be seen from the figures that a decreasing parameter $\alpha$ (from 1 to -8) presses the upper frequency for negative $\beta$ under the upper frequency with positive $\beta$. It is also clear that the $\alpha$ parameter best fits the experimental data with a non-zero $\beta$ parameter, similarly to case of Einstein's gravity. The fact that the best fits are obtained for negative $\alpha$ suggests that the 4D EGB theory is not better suited than Einstein's theory to explain the phenomena.

\section{Conclusions}
\label{Sec:conclusion}

We considered a black hole solution in the $4D$ limit of Einstein-Gauss-Bonnet theory proposed in \cite{Glavan20prl} and studied the motion of charged particles in the spacetime when an external magnetic field is present. The solution reduces to the Schwarzschild case when the GB coupling vanishes. We found the effect of the Gauss-Bonnet coupling on the ISCO for circular orbits of charged particles and determined the collision energy in the vicinity of the horizon. We showed that, similarly to Einstein's case, the presence of the external magnetic field could help explain the observed resonance of high frequency QPOs from microquasars. Also we showed that a better fit of the data is obtained for negative values of the GB coupling constant. However, $\alpha<0$ appears to be disfavoured as it makes EGB gravity repulsive and the black hole can not form dynamically from collapse in this case. While more investigation is needed in order to asses the validity of 4D EGB black holes as possible alternatives to black holes in GR, our study suggests that both classes of solutions show qualitatively similar behaviors and therefore the 4D EGB black holes may not be ruled out solely on the basis of present observations.

\section*{Acknowledgments}
The authors wish to thank A. Aliev for useful discussions and comments. B.A. and S.S. acknowledge Nazarbayev University, Nur-Sultan, Kazakhstan for warm hospitality. This research is supported in part by Projects No. VA-FA-F-2-008 and No. MRB-AN-2019-29 of the Uzbekistan Ministry for Innovative Development and by the Abdus Salam International Centre for Theoretical Physics under the Grant No. OEA-NT-01. D.M. acknowledges support from Nazarbayev University Faculty Development Competitive Research Grant No. 090118FD5348.


\bibliographystyle{elsarticle-num}  
\bibliography{gravreferences}

\end{document}